\documentclass[onecolumn,superscriptaddress,aps,pra,showpacs,floatfix]{revtex4}

\usepackage{latexsym}
\usepackage{amsmath}
\usepackage{amssymb}
\usepackage{latexsym}
\usepackage{color}
\usepackage{graphicx}
\usepackage{cancel}
\usepackage{bbm}
\usepackage{bm}
\usepackage{bbm}
\usepackage{maybemath}
\usepackage{hyperref}

\newcommand{\ii}{\mathrm{i}}

\begin{document}

% =====================
% Weighing the Neutrino
% =====================
\title{Weighing the Neutrino}

\author{U. D. Jentschura}

\affiliation{MTA--DE Particle Physics Research Group,
P.O.Box 51, H-4001 Debrecen, Hungary}

\affiliation{Department of Physics,
Missouri University of Science and Technology,
Rolla, Missouri 65409, USA}

\author{D. Horv\'ath}

\affiliation{Institute for Particle and Nuclear Physics, Wigner
Research Centre for Physics, Hungarian Academy of Sciences, Hungary}

\affiliation{Institute of Nuclear Research of the Hungarian
Academy of Sciences, Hungary}

\author{S. Nagy}

\affiliation{Department of Physics,
Missouri University of Science and Technology,
Rolla, Missouri 65409, USA}

\affiliation{Institute of Physics, University of Debrecen,
H-4010 Debrecen P.O.Box 105, Hungary}

\author{I. N\'andori}

\affiliation{MTA--DE Particle Physics Research Group,
P.O.Box 51, H-4001 Debrecen, Hungary}

\affiliation{Institute of Nuclear Research of the Hungarian
Academy of Sciences, Hungary}

\author{Z. Tr\'ocs\'anyi}

\affiliation{MTA--DE Particle Physics Research Group,
P.O.Box 51, H-4001 Debrecen, Hungary}

\affiliation{Institute of Physics, University of Debrecen,
H-4010 Debrecen P.O.Box 105, Hungary}

\author{B. {U}jv\'{a}ri}

\affiliation{Institute of Physics, University of Debrecen,
H-4010 Debrecen P.O.Box 105, Hungary}

\begin{abstract}
We investigate the potential of short-baseline experiments in
order to measure the dispersion relation of the (muon) neutrino, with a
prospect of eventually measuring the neutrino mass. As a byproduct, the
experiment would help to constrain parameters of Lorentz-violating
effects in the neutrino sector. The potential of a high-flux
laser-accelerated proton beam (e.g., at the upcoming ELI facility),
incident on a thick target composed of a light element to produce
pions, with a subsequent decay to muons and muon-neutrinos, is
discussed. We find a possibility for a muon neutrino mass measurement
of unprecedented accuracy.
\end{abstract}


\maketitle

%
% Introduction
%
\section{Introduction}

The measurement of the neutrino mass is a fundamental, unsolved problem in
elementary particle physics. Indeed, the neutrino is the only particle in the
Standard Model whose basic symmetry property under charge conjugation is still
unknown: If a neutrino were a Majorana particle then it would be equal to its own
antiparticle and described by the Majorana equation, which is manifestly
different from the Dirac equation.  Here, we present a proposal, based on a
specific beamline design, for the measurement of the mass of the muon neutrino
flavor eigenstate, henceforth referred to as the muon neutrino mass.  Our
proposal relies on a short baseline of 1--10\,km, which simplifies the distance
and time measurements as compared to long base-line experiments.  In this
paper, we aim to present a sketch of the beamline design that would be required
to improve current limits on the neutrino mass, as well as a concise discussion
of the theoretical questions and concepts which could be elucidated using the
apparatus.  Indeed, our proposal first and foremost is based on an improved
measurement of the neutrino time of flight (ToF), involving the use of a laser
pulse as a reference signal, but its final aim is to convert the ToF
measurement into a measurement of the neutrino mass.

In the original Standard Model, all neutrinos were assumed to be massless
particles (Weyl fermions), transforming according to the fundamental $(\tfrac12,
0)$ and $(0, \tfrac12)$ representations of the Lorentz group. Neutrino
oscillations were first observed in the deviations of the ratio of solar muon
and electron neutrino fluxes from theoretical predictions.
Laboratory-based studies confirmed the solar 
data~\cite{FuEtAl1998,AhEtAl2002a,AdEtAl2011a,AAEtAl2007}
and fundamentally altered the physical
picture: neutrinos cannot be massless. Moreover, their mass eigenstates cannot
be equal to their flavor eigenstates.  In general, one assumes that in nuclear
reactions neutrinos are originally produced in flavor eigenstates that are
superpositions of mass eigenstates. The mixing is described by the
Pontecorvo--Maki--Nakagawa--Sakata (PMNS) matrix $U$ (for a comprehensive
overview, see Ref.~\cite{StVi2006}). Neutrino oscillations then occur between
the different mass eigenstate components of the original flavor eigenstate. The
measurement of neutrino oscillations has led to rather accurate
measurements of the differences of squared masses of the neutrino mass
eigenstates.  

The absolute values of the neutrino masses however,
in the generalized Standard Model remain entirely unknown.
Their determination requires additional information. Such
independent information can be provided by measuring the
``average specific flavor neutrino mass'' $m(\nu_f)$, defined by
\begin{equation}
m^2(\nu_f) = \sum_{i=1}^3 |U_{f\,i}|^2 \; m_i^2\,,
\end{equation}
where $m_i$ denotes the mass of the mass eigenstates \cite{OtWe2008}.
Determination of the neutrino mass based on the dispersion relation,
such as inferring the mass from ToF measurement, gives information on
$m(\nu_f)$. In the following, the term ``muon neutrino mass'' 
denotes the mass of the flavor eigenstate $m(\nu_\mu)$.

The origin of neutrino masses is still somewhat mysterious.  In the
case of the Cabibbo--Kobayashi--Maskawa (CKM) matrix of the quarks, one
transforms color eigenstates (of the strong interaction) into flavor
eigenstates (of the weak interaction).  In the case of the PMNS-matrix,
one transforms neutrino mass eigenstates into flavor eigenstates (of
the weak interaction), and one might ask if the former might in fact
participate in hitherto unknown interactions
(in addition to gravity). In other words, it would
be somewhat strange if the mass eigenstates have no further physical
interpretation. 

Neutrinos are very special fermions. They are lighter by orders of magnitude as
compared to other ``light'' fermions and therefore approach the light cone
already at very small kinetic energies.  Neutrinos have never been observed at
rest, and in view of the smallness of their low-energy cross sections, it may
never be possible to observe them at rest even if they have nonzero Dirac
and/or Majorana masses. Furthermore, the numerical value of their masses and
moreover, even the sign of their mass squares are not known.  In light of these
experimental results and observations, it is necessary to recall that the best
current limit on the muon neutrino mass $m(\nu_\mu)$ stems from the
investigation of Ref.~\cite{AnEtAl2006}, which puts $m(\nu_\mu) < 2.2 \,
{\rm MeV}/c^2$ at a $90\,\%$ confidence level.  Cosmological observations
provide more stringent limits of the order of 1\,eV for the sum of the masses
of the three neutrino species $\sum_i m_i$ (see
Ref.~\cite{AdEtAl2013planck}), but the interpretation of those measurements
is model dependent \cite{OtWe2008}.  According to Eqs.~(33)--(36) of
Ref.~\cite{AsEtAl1996}, one obtains $m^2(\nu_\mu) = (-0.016 \pm 0.023) \,
{\rm MeV}^2/c^4$ if one uses the value $m_{\pi^-} = (139.569\,95 \pm 0.00035)
\, {\rm MeV}/c^2$ for the negative pion mass, which is close to the preferred
value in the Particle Data Listings \cite{NaPe2012}. 

As mentioned, our proposed measurement of mass is to be based on measuring the
time of flight of muon neutrinos along a short baseline.  The European Light
Infrastructure (ELI) intends to operate high-intensity beamlines used for
proton acceleration, with the intent of answering questions in laser and
particle physics, as well as astrophysics~\cite{ELI_accel,ELI_theor}.  Laser
acceleration of protons with the purpose of generating pions that decay into
muons and muon neutrinos, for injection into a neutrino beamline after passing
through a graphite target, is explicitly mentioned in
Ref.~\cite{BuEtAl2005}, and fundamental particle physics experiments with
laser-driven particle injection is a definite option for the ELI beamlines
which are currently under construction~\cite{Ko2013priv}.  Pions are well known
to be produced copiously when a proton beam is incident on a target composed of
basically any element.  A graphite target would then be a source of a beam of
muon neutrinos. Their energy distribution can be assessed by deflecting the
charged particles out of the line immediately after the production target has
been hit with the laser pulse, thereby eliminating positional uncertainty
regarding the emission of the muon neutrino. After a flight along a short
baseline (of the order of a few kilometers), an obvious approach would be to
detect muon neutrinos in a liquid scintillator detector. 

At high energies, in view of Lorentz invariance, all particles in the
Standard Model approach the light-cone dispersion relation between
energy $E$ and spatial momentum $\vec p$ of the particle,
$E \approx c \, |\vec p|$.  The mass of a particle therefore is
primarily visible in its low-energy properties. Thus in order to assess the
particle's mass, one needs detectors with low energy thresholds. 
Organic liquid scintillators are employed in the Borexino neutrino
detector which is currently in operation at the Laboratorio Nazionale
del Gran Sasso (LNGS). It has a low energy threshold of only $E_\nu =
0.2 \, {\rm MeV}$ (see Ref.~\cite{Le2009}), and is thus able to detect
the ${}^7$Be solar neutrinos which have an energy of about 0.862\,MeV.

A neutrino time-of-flight
measurement could yield a value of the neutrino mass $m(\nu)$, in view
of the relation $E_\nu=m(\nu) c^2/\sqrt{1-v^2/c^2}$, provided the
detector records the energy of the oncoming neutrino and its time of
flight with sufficient precision. 

% For completeness,
% we also mention the dispersion relation of a conceivable superluminal
% neutrino~\cite{ChHaKo1985} which reads as $E_\nu = m(\nu) c^2/\sqrt{v^2/c^2-1}$.

Our paper is organized as follows:
In Sec.~\ref{sec2}, we first provide a mini-review of the status of
theoretical aspects of neutrino physics relevant to our studies. 
Models based on generalized Dirac equations include Lorentz-violation
as a hypothesis. Also neutrino-matter interactions and other
alternative models are discussed.  The experimental proposal is
described in Sec.~\ref{sec31}, while an estimate of the achievable
experimental accuracy is given in Sec.~\ref{sec32}.  Our conclusions
are drawn in Sec.~\ref{sec4}. Finally, in Appendix~\ref{appa}, we briefly
discuss the status of exotic neutrino models with regard to 
fundamental symmetries and the causality principle.  Natural units with
$\hbar=c=\epsilon_0=1$ are used throughout the article unless stated otherwise.

%
% Background Information
%
\section{Theoretical status}
\label{sec2}

In the particle physics data summary of the Particle Data Group
(PDG)~\cite{NaPe2012}, it is stated that ``on the basis of the existing
neutrino data it is impossible to determine whether the massive neutrinos are
Dirac or Majorana fermions''.  Majorana fermions would allow for neutrinoless
double beta decay~\cite{Ro2012}. They are equal to their own
antiparticles~\cite{Pa2010} and have very special transformation properties
under the discrete symmetries. For instance, they have an internal parity equal
to the imaginary unit. This is because in general, we have $P^2 = (-1)^F$ where
$P$ is the parity and $F$ is the fermion number ($F=1$ for Majorana fermions).
The concept of lepton number becomes void if we assume a Majorana nature of the
neutrinos.  The global symmetry of the Standard model Lagrangian under a global
phase change $\psi \to \psi \, \exp(\ii \, \Lambda)$ for all fermion fields
(which otherwise leads to lepton number conservation) is lost if we assume that
the neutrinos are Majorana fermions.  Indeed, they are described by the
Majorana equation which, in four-component notation, is equivalent to the Dirac
equation, supplemented by the charge-conjugation invariance condition. The
Majorana equation, on the level of first quantization, does not allow for
plane-wave solutions proportional to $\exp(-\ii k \cdot x)$, although rather
well-known Dirac-Majorana confusion theorems~\cite{KaSh1981,LiWi1982,Ka1982}
ensure that one cannot reliably distinguish the nature of the neutrino(s) based
on scattering experiments alone.

Right-handed neutrinos and left-handed antineutrinos have never been observed
in nature.  The seesaw (type I, II and~III) mechanisms~\cite{PaSa1974,StVi2006}
explain the lack of observations by the (very) large mass of the ``wrong''
helicity states, which would be of the order of the Grand Unification energy
scale. 

Because of their elusive nature, neutrinos have given rise to exotic
speculation regarding the relativistic wave equation governing their quantum
dynamics.  Indeed, Chodos, Hauser and Kostelecky first
proposed~\cite{ChHaKo1985} that the neutrino might be a tachyon, fulfilling a
Lorentz-invariant dispersion relation while remaining superluminal along its
entire trajectory. Partially inspired by the smallness of the additive terms
needed to give rise to an apparent infinitesimal superluminality of the
neutrino, Kostelecky and collaborators went on to study Lorentz-violating
phases in Standard Model Extensions (SMEs) (see
Refs.~\cite{CoKo1997,CoKo1998,KoMe2012,DiKoMe2009}).  Lorentz-violating
phases would single out a specific direction in space as a ``preferred''
symmetry-breaking seed and lead to anisotropies in the dispersion relation of
particles, thus (potentially) leading to infinitesimal superluminality in
preferred directions.  Lorentz symmetry is not gauged (unless one gauges
gravity).

While Lorentz symmetry is commonly accepted as a basic principle
underlying particle physics, nevertheless in the past numerous works
raised the the possibility of violation of Lorentz invariance
\cite{CoKo1997,CoKo1998,KoMe2012,DiKoMe2009}.  According to the
classic work of Coleman and Glashow~\cite{Coleman:1998ti} one can
introduce Lorentz invariance violating (LV) terms into the Lagrangian,
parameterized by LV parameters. 
LV terms in the Lagrangian can arise due to gravitational, dark energy
or dark matter
interactions~\cite{Cacciapaglia:2011ax,AmelinoCamelia:2011dx}.

In the work of Colladay and Kostelecky~\cite{CoKo1997}, the authors
consider a modified Lagrangian for the Dirac particle as follows:
\begin{equation}
\label{L1}
{\mathcal L} = 
\overline \psi \, \left( \ii \, \gamma^\mu \, \partial_\mu - 
a_\mu \, \gamma^\mu - b_\mu \, \gamma^5 \, \gamma^\mu - m 
\right) \, \psi\,,
\end{equation}
where $\psi$ is the Dirac spinor wave function, and $a_\mu$ and $b_\mu$ are
four-vectors.  Furthermore, $\overline \psi = \psi^+ \, \gamma^0$ is
the Dirac adjoint spinor.
The Dirac equation derived from the
Lagrangian~\eqref{L1} reads
\begin{equation}
\label{KOSDIRAC}
\left( \ii \, \gamma^\mu \, \partial_\mu -
a_\mu \, \gamma^\mu - b_\mu \, \gamma^5 \, \gamma^\mu - m 
\right) \, \psi(x) = 0 \,,
\end{equation}
where $\psi(x)$ is the bispinor Dirac wave function.
Modified Dirac equations of the functional form~\eqref{L1}
do not suppress the ``wrong'' helicity state
of the neutrino field but allow for both 
left-handed as well as right-handed neutrinos;
the additional projections onto the left-handed sector of the 
fermion fields are described in Sec.~3 of Ref.~\cite{KoMe2012}.

Equation~\eqref{L1} breaks Lorentz
invariance, but in a very subtle way. Namely, under a Lorentz transformation
the Dirac field transforms as $\psi'(x') = S(\Lambda) \, \psi(x)$,
where $S(\Lambda)$ is the bispinor representation of the Lorentz group.
We recall the well-known relation
\begin{equation}
\gamma'^\mu = {\Lambda^\mu}_\nu  \, S(\Lambda) \,
\gamma^\nu \, S(\Lambda)^{-1} = \gamma^\mu \,.
\end{equation}
This implies that under the combined action of the spinor and vector 
Lorentz transformations, the Dirac matrices are shape-invariant.
Therefore, upon a change of the Lorentz frame, the Dirac 
equation changes to
\begin{equation}
\label{KOSDIRAC2}
\left( \ii \, \gamma^\mu \, \partial'_\mu -
a'_\mu \, \gamma^\mu - b'_\mu \, \gamma^5 \, \gamma^\mu - m
\right) \, \psi = 0 \,,
\end{equation}
For zero $b_\mu$, the dispersion relation fulfilled 
by the solutions of Eq.~\eqref{KOSDIRAC} reads as 
\begin{equation}
(E -a_0)^2 - (\vec p - \vec a)^2 = m^2 \,,
\qquad b_\mu = 0 \,,
\end{equation}
whereas the dispersion relation for Eq.~\eqref{KOSDIRAC2} reads as
\begin{equation}
(E' -a'_0)^2 - (\vec p' - \vec a')^2 = m^2 \,,
\qquad b'_\mu = 0 \,.
\end{equation}
Here, $\vec a = (a^1, a^2, a^3)$ denotes the spatial part
of the four-vector $a^\mu$. 
If the $a_\mu$ and $b_\mu$ were ordinary Lorentz 
vectors, then Lorentz invariance would be fulfilled. 
However, it is stressed in Ref.~\cite{CoKo1997} that,
in keeping with their interpretation as effective couplings arising
from a scenario with spontaneous symmetry breaking (of Lorentz
symmetry), $a^\mu$ and $b^\mu$ are invariant under CPT transformations.
In the text before Eq.~(5) of Ref.~\cite{CoKo1997}, it is also
stressed that (because the currents $\gamma^\mu$ and $\gamma^5 \,
\gamma^\mu$ change sign under CPT), the Lorentz-breaking terms are not
CPT invariant. 
Any experiment which investigates Lorentz invariance in
the neutrino sector constrains the parameters $a^\mu$ and $b^\mu$.

A conceivable relevance of Eq.~\eqref{L1} 
for neutrino physics can be tested experimentally,
and it even allows for a connection to be drawn
to the original work (Ref.~\cite{ChHaKo1985}) in 
which an exotic variant of the Dirac equation was proposed 
for neutrino physics. Namely, if we
study Eq.~\eqref{L1} in a Lorentz frame with $b^\mu = 0$ and
$\vec a = \vec 0$, then the generalized Dirac Hamiltonian reads as
$\vec\alpha \cdot \vec p + a_0 + \beta \, m$
($\vec\alpha = \gamma^0 \, \vec\gamma$ and $\beta$ matrices 
are used in the Dirac representation).
For $m=0$ and negative $a_0$, i.e., $a_0 = -|a_0|$, it reads as
\begin{equation}
H = \vec\alpha \cdot \vec p - | a_0 | \qquad \to \qquad
E = |\vec p| - | a_0 | \,.
\end{equation}
If an experiment is restricted to a specific 
energy range and cannot scan the dispersion relation,
then it is instructive to point out that the tachyonic dispersion
relation~\cite{ChHaKo1985} $E = \sqrt{\vec p^2 - m^2}$
approximates the dispersion relation of the
Lorentz-violating Dirac equation with constant $a_0$ 
in a specific energy interval
($E, |\vec p| \gg m$),
\begin{equation}
E = \sqrt{ \vec p^{\,2} - m^2 } \approx
|\vec p| - \frac{m^2}{2 \, |\vec p|} \approx
|\vec p| - | a_0 | \,,
\qquad
|a_0| \approx \frac{m^2}{2 \, |\vec p|} \approx \frac{m^2}{2 \, E} \,.
\end{equation}
In this case, one would intuitively assume that 
spatial anisotropies modeled by other nonvanishing $a_\mu$ and
$b_\mu$ parameters and measurements over wider energy ranges
would allow us to constrain the parameter space and the 
functional form of the Lorentz violation further.
Also, one should note that constant shifts on the neutrino energy (such
as the one produced by $|a_0|$) cannot be determined by a time-of-flight
measurement because the group velocity remains unchanged. 
In order to produce an observable effect, an additional
modification of the dispersion relation due to 
a weak decay is necessary, as indicated in Ref.~\cite{DiKoLe2013}.

Other recently explored models describe a
conceivable environmental effect on neutrino propagation
\cite{Dvali:2011mn,Ke2011cb,Oda:2011kh,Iorio:2011ay}. 
One can write down Lagrangians where a background-dependent fifth-force
field couples to neutrinos differently as compared to other particles.
The fifth-force interaction might be enhanced in spatial regions with a
high matter density. According to Ref.~\cite{Dvali:2011mn}, one may
speculate that the interaction between the new field and the neutrino
sector could modify the effective metric seen by the neutrinos which
could result in modifications of the ``local'' dispersion relation
without violating the Lorentz symmetry at a fundamental level. In
Ref.~\cite{Dvali:2011mn}, the authors introduce a new massive spin-2
field in order to formulate their model. By contrast, in
Ref.~\cite{Oda:2011kh}, Oda and Taira formulate a model in terms of a
spin-1 gauge field sourced in a condensed-matter environment.

It is claimed in recent papers~\cite{Eh2013,ChEh2013} that neutrinos
from the supernova SN1987A data support the existence of 4.0\,eV and
21.4\,eV neutrino mass (not flavor!) eigenstates, and it is shown that
such large masses could be made consistent with existing constraints
including neutrino oscillation data and upper limits on the neutrino
flavor state masses, provided that there also exist a pair of sterile
neutrino mass states whose masses are nearly degenerate with the active
ones, plus a third active-sterile doublet that is tachyonic ($m^2 <
0$).  The most direct confirmation might involve a neutrino oscillation
experiment sensitive to $\Delta m^2=(21.4)^2-(4.0)^2 = 442\,{\rm eV^2}$. 
For example, at a neutrino energy of 1\,GeV one easily finds an
oscillation wavelength of about 34\,cm, which could be readily observed
in a short-baseline experiment, realized with a spatially compact
experimental apparatus to be described in the following.

At the end of the current section, 
we reemphasize that of course,
all of the exotic theoretical model discussed above 
remain highly controversial (see also Appendix~\ref{appa}).

%
% Proposition: test of neutrino speed at a shorter distance
% 
\section{Experimental proposal}
\label{sec3}

%
% Beamline Proposal
%
\subsection{Beamline setup}
\label{sec31}

\begin{figure}[t!]
\begin{center}
\begin{minipage}{0.9\linewidth}
\begin{center}
\includegraphics[width=1.0\linewidth]{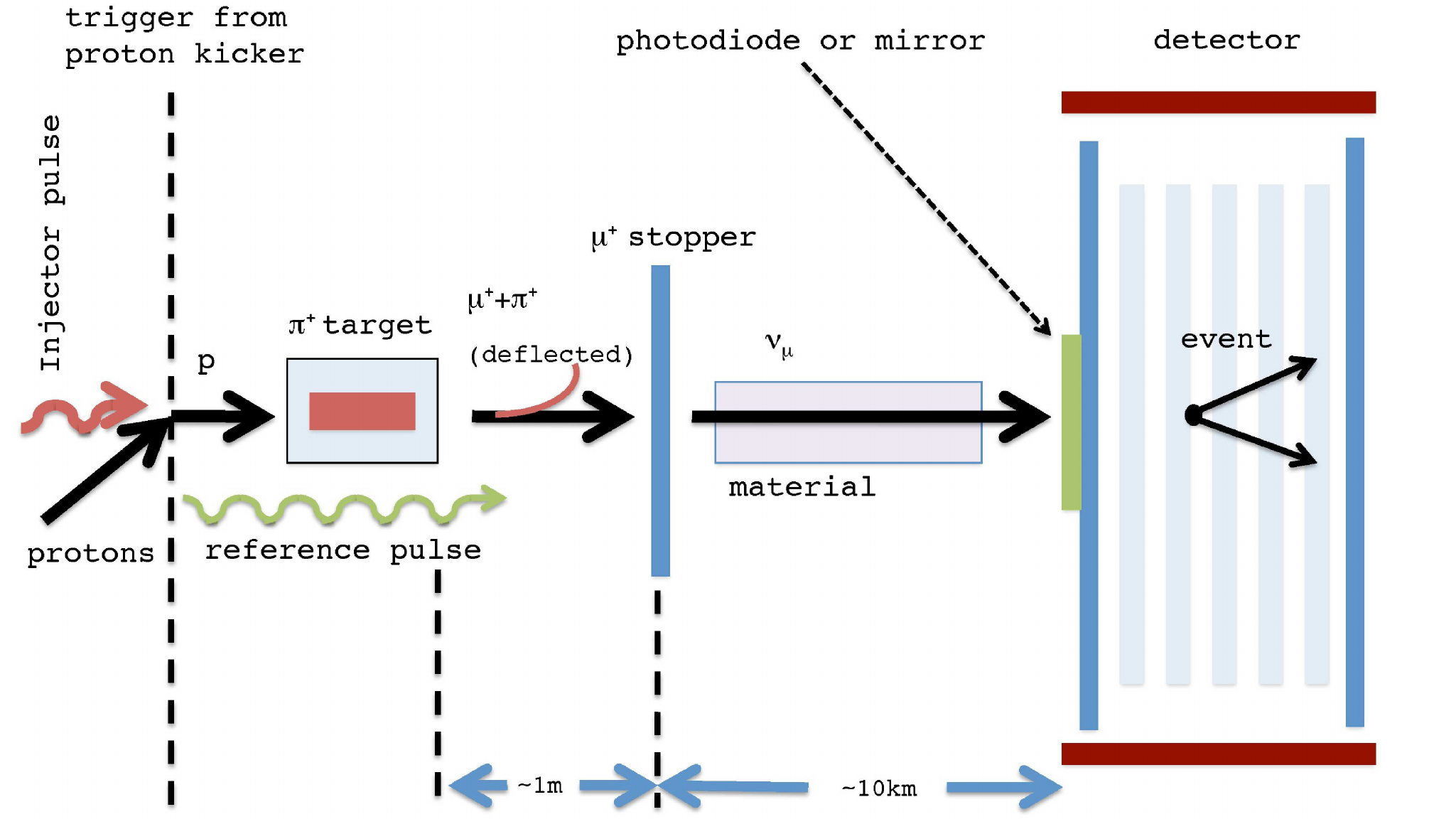}
\caption{\label{fig1} Outline of the proposed experiment.
The explanation is in the text.}
\end{center}
\end{minipage}
\end{center}
\end{figure}

Our proposal for the beamline setup is sketched in Fig.~\ref{fig1}. 
Protons can either be fed into the beam line from a storage ring, via a
kicker magnet (as shown in the figure), or they can be injected into
the beam line by laser acceleration~\cite{BuEtAl2005} (``injector
pulse''). Indeed, the idea of injecting protons into a neutrino
experiment by laser acceleration was formulated in
Ref.~\cite{BuEtAl2005}.  With current technology, it is possible to
reach laser intensities of $10^{22} \, {\rm W}/{\rm cm}^2$ (see
Ref.~\cite{BaEtAl2004}), which is more than sufficient to immediately
accelerate a ``plasma mirror'' consisting of protons and electrons into
the relativistic regime.  As discussed in Ref.~\cite{BuEtAl2005},
during the laser interaction the radiation momentum is mainly
transferred to the protons through the charge separation field (in the
plasma). The kinetic energy of the protons therefore is much greater
than that of the electrons.

Our proposal aims at comparing directly the speeds of neutrinos and
photons, in a relatively short baseline, as given in Fig.~\ref{fig1}. 
One may envisage to send photons either together with the neutrinos in the
same vacuum pipe, or in a tiny vacuum pipe in the middle of the
neutrino ``tunnel''. The reference laser pulse is
%(``reference pulse'', different from the proton injector pulse) needs to be
triggered by the
proton injector device, with a constant time delay.  The short
baseline could also facilitate a neutrino speed measurement performed
in vacuum and in baryonic matter simultaneously, in order to test a
conceivable matter dependence.

In the (graphite) neutrino production target, the protons generate a
highly collimated beam of positively charged pions that can decay into
positively charged muons ($\mu^+$) and muon neutrinos ($\nu_\mu$). 
Muons and pions are deflected from the beam line almost immediately
behind the pion target, in order to restrict the spatial region along
the beam line available for muon neutrino creation. For typical muons
and pions produced in p--N collisions, the relativistic factor is of
order unity, and a typical value is $\gamma \approx 2$.
The masses of muon and pion are $m \approx 106$ and $\approx 139$\,MeV,
and the magnetic field required to deflect these charged particles
out of the beam on a distance of $R \approx 10\,$cm is
given by the cyclotron formula
\begin{equation}
B = \frac{\gamma \, m \, v}{e R}  \approx 8 \, {\rm T}\,,
\end{equation}
which is of the same order-of-magnitude as currently used for 
high-performance magnets such as those used at the 
Large Hadron Collider (LHC). 
The finite lifetime of the $\pi^+$ decay into $\mu^+ + \nu_\mu$
otherwise leads to an ambiguity in the time of flight of the decay
products, which we would like to eliminate as far as possible,
by restricting the spatial region in which the pions may decay.  Still,
in order to eliminate a conceivable admixture from the high-energy tail
of laser-accelerated protons (and high-energy pions/muons), the
insertion of a muon stopper into the beamline appears useful.  This
approach necessitates a high-intensity proton source, which can be
realized with a strong laser.  Let us assume that, with a pion target
of dimension 5\,cm along the beam line, muons and pions are allowed to
travel another 10\,cm before being deflected. This takes roughly
$3 \times 10^{-10}$\,s at the speed of light, resulting in a comparable
uncertainty of order $10^{-10}$\,s as to the point where the neutrino 
creation occurred.

Converted into a time-of-flight measurement, the ensuing uncertainty in
the assignment of the time of flight of the neutrino (as opposed to
that of the pion) leads to an experimental ``line shape'' 
amenable to a simulation 
with Monte Carlo techniques (GEANT4, see Ref.~\cite{GEANT4}), given
the concrete experimental parameters and details of the apparatus. 
An estimate of the experimental beam shape at the detector based
on GEANT4 is included below.
Line shapes can typically be analyzed to one part in 100, so that
optimistically, a projected resolution on the order of $3\times
10^{-12}$\,s for the time-of-flight measurement seems within reach.
The lifetime of a pion is on the order of $2.6 \times 10^{-8}$\,s.  At
a typical relativistic Lorentz factor of $\gamma = 2$, i.e., for
$280\,$MeV~pions, we can thus expect that in the spatial region before
deflection about half a percent of the created pions will in fact
decay into a muon-neutrino, the ratio $\chi$ of the decaying
pions (in the restricted spatial region) to the overall number of 
pions being given by
\begin{equation}
\chi \approx \frac{3 \times 10^{-10} \, {\rm s}}%
{2 \times \left( 2.6 \times 10^{-8} \, {\rm s} \right)} = 0.0057 \,.
\end{equation}
With the copious production of pions taken into account, this should
not inhibit the possibility of obtaining good statistics in the
measurement.

The neutrinos then continue their flight along the beam axis for a distance of
the order of 10\,km. We found this length scale as a good compromise in order
to make the observation of a sizeable deviation from the speed of light
possible with typical currently available detector time resolution, while
keeping the construction costs under control and allowing for a reference light
pulse to be sent from the graphite target to the detector, eliminating any
uncertainties regarding the actual length of the beamline (see also
Fig.~\ref{fig1}). In order to make the apparatus operational for higher-energy
(above 50\,MeV) neutrinos, one has to surround the primary detector with a
``veto'' detector which reveals when a cosmic ray passes into the apparatus,
allowing the corresponding activity in the primary detector to be ignored
(``vetoed'').  For low-energy experiments, one typically has to locate the
detector deep underground so that the earth above could reduce the cosmic ray
rate to tolerable levels, indicated by horizontal bars around the detector in
Fig.~\ref{fig1}.

With a time tagging of the order of a few picoseconds available in
typical modern detectors, it should then be possible to derive
meaningful information on the neutrino dispersion relation.  However, a
critical question concerns the basic properties of the neutrino
detector suitable for such an experiment. Both
Charged-current interaction as well as 
neutral-current interactions are relevant to the
detection of muon neutrinos, as given by the following 
Feynman diagrams (a) and (b)
\begin{center}
\includegraphics[width=0.6\linewidth]{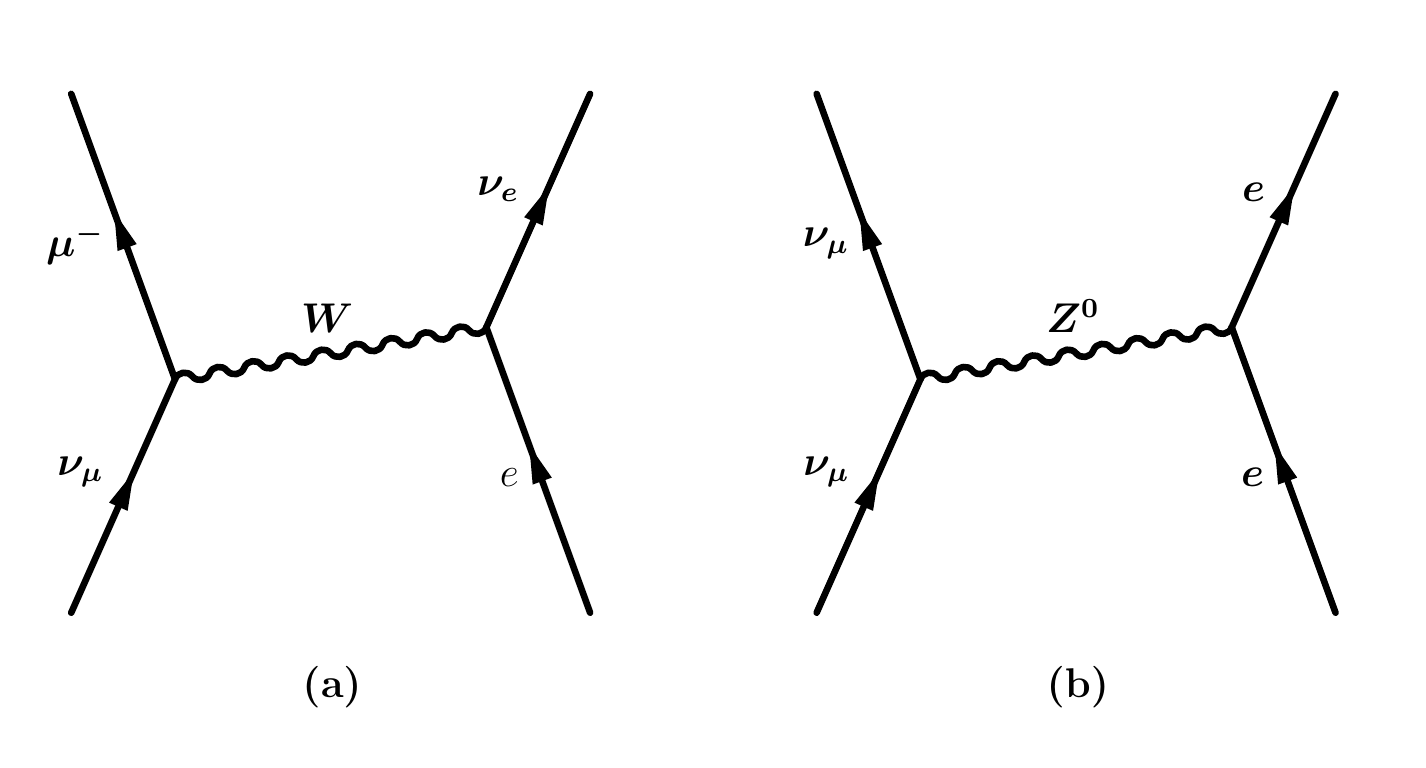}
\end{center}
respectively. Above
muon threshold, weak charged-current (CC) interactions $\nu_\mu + N =
\mu^- + X$ (where a nucleus in the material undergoes a transition  $N
\to X$), or $\nu_\mu + e^- = \mu^- + \nu_e$ (with the electrons in the
detector material), produce muons which can be detected, e.g., via
their Cherenkov light in water (as in the Super-Kamiokande
detector).  For muon neutrinos, the threshold is about 105 MeV.
For lower muon neutrino energy, charged-current interactions are not
available in order to trigger detection. Neutral-current interactions
(like elastic scattering) take over at lower energies.  

Indeed, for low-energy neutrinos, liquid scintillator detectors are
much better, because the interaction of a neutrino (of any flavor) via
neutral-current interactions with the electrons in the scintillator
medium (scattering process $\nu_\mu + e \to \nu_\mu + e$) does not have
any threshold at all.  Indeed, the MiniBooNE detector
\cite{StEtAl2001,AAEtAl2007,AAEtAl2008,AAEtAl2009prl1,AAEtAl2009prl2,AAEtAl2010}
uses pure mineral oil as a detection medium, which is a natural
scintillator.  In order to illustrate this statement, we recall that
the MiniBooNE detector is filled with about 800~tons of mineral oil,
and its 1280 photomultipliers provide coverage for about 10\,\% of the
tank region.  No reliable information seems to be available on the
precise value of the MiniBooNE detector threshold, but data taken for
Ref.~\cite{AAEtAl2009prl2} (available on the webpage of the
experiment, see Ref.~\cite{MiniBooNEDataRelease})
go down to $200\,{\rm MeV}$,
suggesting that the detector threshold must be around this value or below.

The Borexino detector uses 
300 tons of organic liquid scintillator pseudocumene (PC) containing
1.5\,g PPO in each liter of PC, immersed in 1000~tons of pure PC as buffer.
The buffer is surrounded by 2200 photomultipliers. An external shielding
is provided by 2400 tons of pure water. Its threshold~\cite{Le2009}
is at roughly $0.2\,{\rm MeV}$, well below the $0.862\,{\rm MeV}$ of
the ${}^7$Be neutrino flux and thus responsive to a very important part
of the solar neutrino spectrum (see Table 6.2 of Ref.~\cite{StVi2006}).
Liquid scintillator detectors typically have a good time and energy
resolution, but do not preserve directional information.  While they
are not as fast as Cerenkov detectors, they currently hold the record
for the lowest energy threshold among neutrino detectors.

\begin{figure}[t!]
\begin{center}
\begin{minipage}{1.0\linewidth}
\begin{center}
\includegraphics[width=0.97\linewidth]{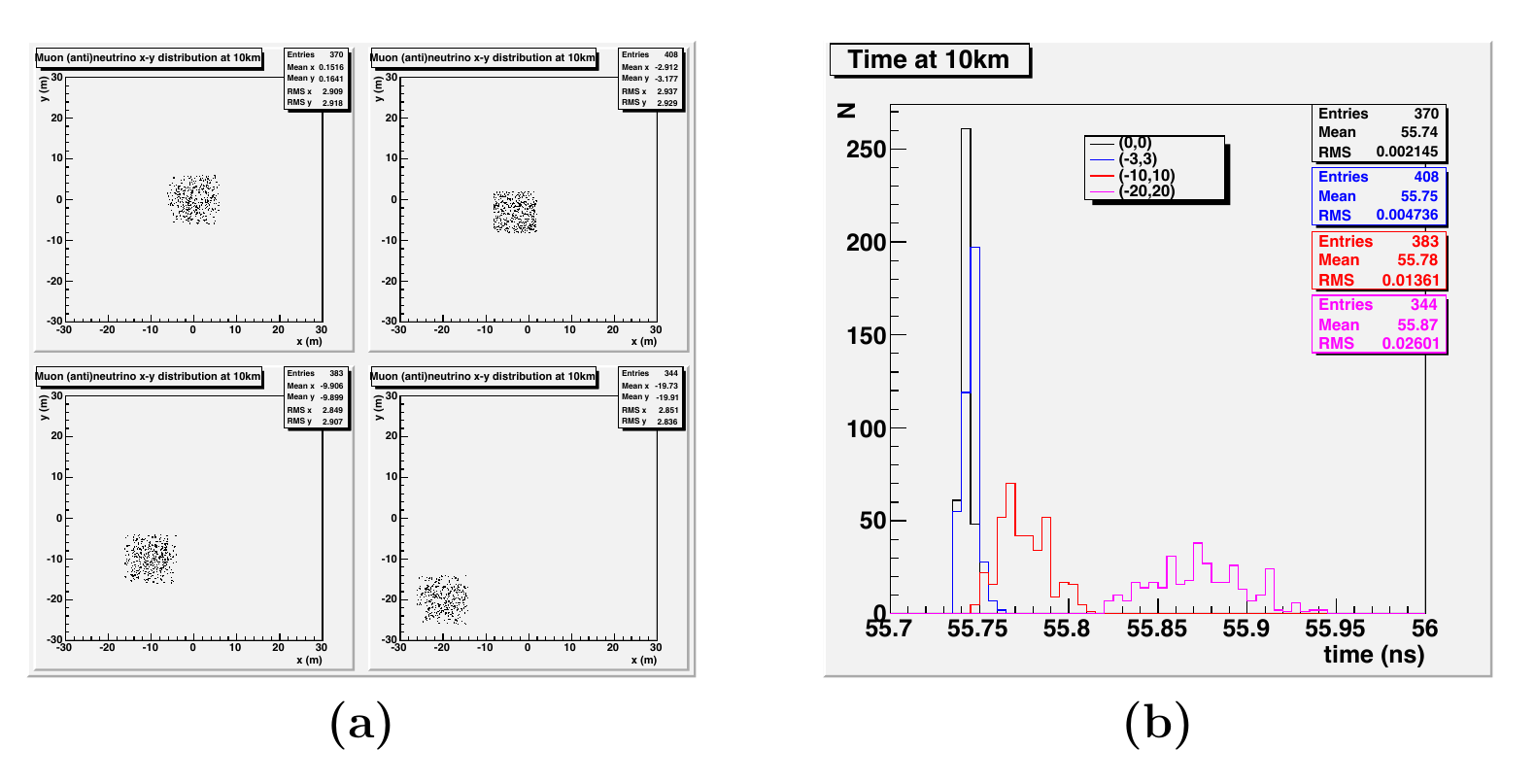}
\caption{\label{fig3}%Rough 
Figure (a) gives an
estimate of the event distribution 10\,km away from the graphite
target at different points in the $xy$ plane,
while Fig.~(b) estimates time distribution of the neutrino ToF delay as
compared to the ToF of light belonging to a 10\,km distance.
See the text for further explanation.}
\end{center}
\end{minipage}
\end{center}
\end{figure}

In order to give some guidance on the detector design, the result of a
GEANT4~Monte-Carlo simulation with 100 billion protons with energy of
$450\, {\rm GeV}$ and positively charged pions in the energy range of
$(100\pm1)$\,GeV are presented in Fig.~\ref{fig3}. The neutrino
distribution resulting from the decay channel $p \to \pi^+ \to \mu^+ +
\nu_\mu$ is evaluated 10\,km away from the graphite target placed on the $z$ axis. 
All of the decays are simulated.  The target is assumed to consist of a
2\,mm thick carbon foil located at $z = 0$.

Along the first 100\,m in $z$, we assume propagation in a vacuum pipe
and we placed rock from $z = 100$\,m to 10\,km. At 10\,km there is a
virtual screen to exhibit the neutrino distribution.  We use neutrinos
from the process $\pi^+ \to \mu+ \nu_\mu$, with subsequent decay of the
anti-muon, $\mu^+ \to e^+ \nu_\ell \overline{\nu}_\mu$. We assume that
there is no chance to distinguish muon neutrino and anti-muon neutrino.
In Fig.~\ref{fig3}(a), the distribution of neutrinos in the
$xy$ plane is shown in a $10\times 10$\,m$^2$ area (the surface of a
detector) with different centres to show the 
distribution of the ToF, starting
at 33.3\,$\mu$s for the shortest path. In Fig.~\ref{fig3}(b), 
we show the positions of the centres  in the $xy$ plane in
meters. The root-mean-squared of the ToF's shows that if the centre of
the detector is close to the z axis, neutrinos cross its surface in a
time interval of a few picoseconds (ps).  A large fraction of pions decay 
into muon and neutrinos in the first few meters of the rock, but the decay products
lose the $z$ direction, and their contribution to the neutrinos at the
detector is negligible.

With an ideal proton bunch and a detector positioned precisely (with an
uncertainty of at most $\pm 10$\,cm in the $x$-$y$ plane) the ToF
differences are few ps.

%
% Estimate of the Experimental Accuracy
%
\subsection{Estimate of the experimental accuracy}
\label{sec32}

Based on the considerations reported in Sec.~\ref{sec31}, we assume
that the time resolution of the detector is about $\delta t \approx 3$\,ps.
We denote the time of flight as $t_0$, the length of the baseline as
$s_0$. The speed of the neutrino is $v_{\nu} =  s_0/t_0 = (1+\delta)\,c$,
so the speed of light is $c  = s_0/t_0 + O(|\delta|)$ 
(with $\delta < 0$).
The smallest measurable deviation from the
speed of light $\delta c$ and the mass of the neutrino $m(\nu)$ are
related as
\begin{equation}
m(\nu) = \frac{E_\nu}{c^2} \, \sqrt{2 \, |\delta|} + O(|\delta|^{3/2}) \,,
\qquad
\qquad
|\delta| = \frac{\delta t}{t_0} \approx c \, \frac{\delta t}{s_0} \,.
\end{equation}
For a baseline of $10\,$km with $|\delta c| \approx (3 \times 10^8)^2
\,(3 \times 10^{-12}) \, 10^{-4} {\rm m}/{\rm s} = 30\,{\rm m}/{\rm s}$,
$|\delta| \approx 10^{-7}$, and with $\sqrt{2 \, |\delta|} = 0.00042$,
assuming a 1\,MeV neutrino, one can determine the neutrino mass to
$420$\,eV, which is an improvement over Ref.~\cite{AnEtAl2006} by (more
than) three orders of magnitude. Based on measured neutrino mass squared
differences~\cite{StVi2006}, one can infer that the heaviest neutrino
cannot be lighter than $0.05\, {\rm eV}$. 

It is interesting to relate this projected accuracy,
assuming spatial isotropy, to the
Lorentz-violating parameters in the formalism of Ref.~\cite{KoMe2012}. 
From Eq.~(83) of Ref.~\cite{KoMe2012} (omitting flavor indices), we
have the Lorentz-violating dispersion relation
\begin{equation}
E = |\vec p| + \frac{m(\nu)^2}{2 |\vec p|} -
|\vec p|^{d-3} \, \left(a^{(d)}_{\rm of} + c^{(d)}_{\rm of} \right) \,,
\end{equation}
where $d$ denotes the dimensionality of the $a$ and $c$ coefficient,
which in turn parameterize the Lorentz-violation in the formalism of
Ref.~\cite{KoMe2012}.  For oscillation-free (flavor-blind) models
(subscript ``of''), the $d$-dimensional S-wave coefficients
$(a^{(d)}_{\rm of})_{00}$ and $(c^{(d)}_{\rm of})_{00}$ can be
determined from the relations [see Eq.~(126) of Ref.~\cite{KoMe2012}],
\begin{subequations}
\begin{align}
\label{eqa}
-(d-3) \, |\vec p|^{d-4} \,
{}_0{\mathcal N}_{00} \;\; (c^{(d)}_{\rm of})_{00} =& \; \delta  \,,
\\[2ex]
\label{eqb}
(d-3) \, |\vec p|^{d-4} \,
{}_0{\mathcal N}_{00} \;\; (a^{(d)}_{\rm of})_{00} =& \; \delta  \,.
\end{align}
\end{subequations}
In the following, we use $d=4$ in the first and $d=5$ in the second
equation.  Here, $\delta c$ is the deviation from the speed of light
and ${}_0{\mathcal N}_{00} = 1/\sqrt{4 \pi} \approx 
0.28$ is a geometric factor entering the leading, isotropic term 
of the spherical decomposition of the Lorentz-violating effects.
This value of ${}_0{\mathcal N}_{00}$ should 
be used for the MINOS, OPERA, and T2K experiments, 
as given by Kostelecky and
Mewes in Table~X of Ref.~\cite{KoMe2012}.

Using $d=4$ in Eq.~\eqref{eqa}, one easily verifies the entry
$(c^{(4)}_{\rm of})_{00} \approx (-8.4 \pm 1.1) \times 10^{-5}$ in
Table XI of~\cite{KoMe2012} which would otherwise correspond to the
retracted OPERA result. From our proposed beamline, one could
potentially derive much stricter bounds, namely, using our projected
accuracy of $\delta c \approx 30 {\rm m}/{\rm s}$, one could 
potentially obtain a bound on the order of
\begin{equation}
\left| (c^{(4)}_{\rm of})_{00} \right| \approx 3.57 \times 10^{-7} \,,
\end{equation}
which is tighter by two orders of magnitude.  For the coefficient
$(a^{(5)}_{\rm of})_{00}$, the bound is less stringent, 
$\left| (a^{(5)}_{\rm of})_{00} \right| \approx 1.78 \times 10^{-4} 
\, {\rm GeV}^{-1}$,
because of the scaling of this effect with the energy.  For stricter
bounds on $(a^{(5)}_{\rm of})_{00}$, one should perform the experiment
at higher energies, which would however contradict the basic notion 
behind our experiment, which is to measure the muon neutrino mass.

We should note that astrophysical bounds on LV parameters are in general
stricter.  For instance, the modulus of the coefficient $(c^{(4)}_{\rm
of})_{00}$ has been constrained at the $10^{-8} \, \ldots \, 10^{-9}$ level
using data from the SN1987A (according to Table~XII of
Ref.~\cite{KoMe2012}). Moreover, using cosmic rays, the upper bound on this
coefficient is of order $10^{-11}$ (see page~4 of Ref.~\cite{Al2009}).  A
lower bound (for negative values, i.e., a bound not excluding zero) is of order
$10^{-19}$ based on IceCube data (see Table II of Ref.~\cite{DiKoMe2013}.
Similarly, the modulus of the coefficient $(a^{(5)}_{\rm of})_{00}$ has been
constrained at the $10^{-7}$ level using data from supernovas~(according to
Refs.~\cite{KoRu2011,KoMe2012}).  Terrestrial bounds on the coefficients,
which are accessible from our experimental proposal, are much more stringent
from the point of view of reproducibility. The full control over the
experimental parameters in a terrestrial experiment would make it possible to
scan a larger momentum or energy interval in the Lorentz-violating terms (which
are proportional to $|\vec p|^{d-3}$).  Regarding bounds for Lorentz-violating
terms derived from the supernova 1987A, one may point out that the derived
bounds  strongly depend on whether one includes the mysterious ``early'' burst
under the Mont Blanc~\cite{BiEtAl1987JETPL} or not.

%
% Conclusions
%
\section{Conclusions}
\label{sec4}

Our proposed experimental apparatus sketched in Fig.~\ref{fig1} has a lot of
parameters which can be adjusted in order to accomplish a meaningful
time-of-flight measurement of muon neutrinos. One has tight control over the
time measurement and does not need GPS calibration as is otherwise necessary
for much longer baselines.  With laser acceleration, or by feeding from a
storage ring, one can taylor the energy of the injected protons and hope to
measure the dispersion relation of the muon neutrinos.  We believe that the
idea of a well-controlled, short or medium-baseline environment (see also
Sec.~6 of Ref.~\cite{AdEtAl2013}) for the measurement of the muon neutrino
mass has some potential for future experiments.

Still, because of the smallness of the neutrino mass, and without considerable
further improvement in neutrino detectors beyond the current state of the art
in liquid scintillators, in a first approximation, neutrino time-of-flight
measurements should be classified as probes of Lorentz violation (as described
in Sec.~\ref{sec2}).  Furthermore, in a short-baseline experiment, ``local''
effects on time-of-flight measurements of neutrinos~\cite{ASEtAl2012} can be
tested in a very controlled manner, because it is possible to insert a specific
material into the beamline and test for matter interaction effects. Long
baselines are not necessarily required in order to test interesting physics
(see Sec.~\ref{sec3}), but rather, one can argue that in scenarios with
comparatively large mass differences among the neutrino mass
eigenstates~\cite{Eh2013}, a short-baseline experiment would be best suited for
a comprehensive study.  This is also the case for seeing a possible tachyonic
neutrino of mass-squared  $-0.2 \, {\rm keV}^2$, which could be observed under
the experimental conditions we propose~\cite{Eh2013}.  A possible beamline 
setup is discussed in Fig.~\ref{fig1}.  With current detector technology, one can
still hope to improve the (limits on the) muon neutrino mass by several orders
of magnitude, and improve the current (terrestrial) value of $m_{\nu_\mu} < 2.2
\, {\rm MeV}$~\cite{AnEtAl2006} into the keV~regime (see Sec.~\ref{sec32}).
Cryogenic bolometers~\cite{Gi2012} may pave the way into the future.

%
% Acknowledgements
%
\section*{Acknowledgements}

%TAMOP 4.2.4.A/2-11-1-2012-0001 

This research was supported by the National Institute of Standards and
Technology (Precision Measurement Grant), the National Science Foundation
(Grant PHY-1068547), the Hungarian Scientific Research Fund grants OTKA
NK81447, K103917 and the TAMOP 4.2.2/B-10/1-2010-0024,
TAMOP-4.2.2.C-11/1/KONV-2012-0010 projects (National Excellence Programs
subsidized by the European Union and co-financed by the European Social Fund).
U.D.J.~thanks the MTA-DE Particle Physics research
at the Academy Institute in Debrecen for kind hospitality during a guest
researcher appointment in summer 2013, during which essential stages of this work were
completed.

\appendix

%
% Sending Information into the Past
%
\section{Exotic Lorentz--Conserving Dirac Equations}
\label{appa}

\subsection{Tachyonic Dirac Equation}

In view of the relevance of neutrino ToF experiments for the exclusion of
exotic theoretical models, we briefly recall in this Appendix the so-called
tachyonic (superluminal) Dirac equation, proposed originally in
Ref.~\cite{ChHaKo1985} and recently subject to a few  more investigations
reported in 
Refs.~\cite{JeWu2012epjc,JeWu2012jpa,JeWu2013nemeti,JeWu2013isrn}.
It is based on the Lagrangian
\begin{equation}
\label{L5}
{\mathcal L} =
\left( \overline \psi \, \gamma^5 \right) \, 
\left( \ii \, \gamma^\mu \, \partial_\mu - \gamma^5 \, m \right) \, 
\psi\,.
\end{equation}
The plane-wave solutions of the variational equation
\begin{equation}
\label{TACDIRAC}
\left( \ii \, \gamma^\mu \, \partial_\mu - \gamma^5 \, m \right) \, \psi(x)
= 0
\end{equation}
fulfill the tachyonic (superluminal) dispersion relation
$E^2 = \sqrt{\vec p^2 - m^2}$.  Quite 
recently~\cite{JeWu2012epjc,JeWu2012jpa,JeWu2013isrn}, 
it has been shown that the solutions of Eq.~\eqref{TACDIRAC} have an
convenient structure in helicity basis~\cite{JeWu2013isrn}. 
Furthermore, they fulfill projector sum rules, discussed in
Refs.~\cite{JeWu2012epjc,JeWu2013isrn}, permitting the evaluation of
the tachyonic Dirac propagator as the inverse of the kinetic operator
(Dirac Hamiltonian), which is crucial
in formulating the corresponding field theory.  In view of the structure of the
Lagrangian~\eqref{L5}, the role of the Dirac adjoint in the tachyonic
sector is assumed by the pseudoscalar adjoint~$\overline \psi\, \gamma^5$,
and the generalized Dirac propagator is calculated as the time-ordered
product $S_T(x,y) = 
\left< 0 \left| T \; \psi(x) \, \left( \overline \psi(y) \gamma^5 \right)
\right| 0 \right>$.
One postulates (anti-)commutator relations for the fundamental field
operators which assign negative norm to the ``wrong'' helicity states
of the tachyonic field, excluding them from the physical spectrum via a
Gupta--Bleuler condition~\cite{JeWu2013isrn}. 
In particular, vacuum instability issues, which plague spin-zero tachyonic
field theory, do not affect the spin-$\tfrac12$ version, because the
(pseudoscalar) mass term enters the Lagrangian linearly, not
quadratically. 

Let us take back a step and assume that neutrinos 
are Dirac particles and travel slower than light.
The elusive helicity reversal of a neutrino 
overtaken on a highway without speed limits 
(in a ``thought experiment'') has given rise to a number of pertinent
investigations and discussions~\cite{Fe1998q76,Ho1998q76,GoGo2011}.
The intriguing helicity reversal has
been discussed as ``question \#{}76'' in Refs.~\cite{Fe1998q76,Ho1998q76}.
If we assume that the neutrino is a subluminal Dirac particle,
then right-handed neutrino interactions are suppressed
in the (ordinary) Dirac theory by a factor $m_\nu/E$
where $E$ is the energy scale,
and one might thus argue that overtaking a Dirac 
neutrino sterilizes it. However, the ``overtaken and now right-handed
neutrinos'' are not completely sterile; their transition
currents are suppressed but not zero
(helicity is not equal to chirality if the mass is nonvanishing).
The ``overtaken neutrinos''
are, in particular, less sterile than the 
``other'' right-handed neutrinos which otherwise arise
from the seesaw mechanism: Interactions of the latter ones
are suppressed by an additional factor $m_\mu/M_{\rm GUT}$
where $M_{\rm GUT}$ is the Grand Unification Scale.
By contrast, the ``overtaken and now right-handed'' subluminal Dirac neutrinos
have the same mass as the left-handed ones by
virtue of Lorentz symmetry conservation during the overtaking maneuvre.
We also recall that an
essential ingredient in the construction of the original Standard Model
was that neutrinos are massless (and
transform as a Weyl spinor, which allows them to be either left- or
right-handed).

%
% Why a Tachyonic Neutrino has to be Light: Causality and Information
%
\subsection{Why a Tachyonic Neutrino has to be Light: Causality and Information}

The mysterious neutrino burst under the Mont Blanc~\cite{BiEtAl1987JETPL},
which was recorded roughly 4 hours before the rest of the other
neutrinos from the supernova SN1987A still eludes (some) scientists.
The classical dispersion relation, or, energy-velocity relation, of a
superluminal particle reads as follows (we revert to SI units),
\begin{equation}
\label{Enu}
E_\nu = \frac{m_\nu \, c^2}{\sqrt{(v/c)^2 - 1}} \approx 
\frac{m_\nu \, c^2}{\sqrt{2}} \, \sqrt{\dfrac{c}{\delta c}} \,,
\qquad
\qquad
v = c + \delta c \,.
\end{equation}
Here, we would like to discuss the question why the exotic
tachyonic described by Eq.~\eqref{TACDIRAC}, if they exist, have to 
be very light particles, and we also address a certain 
caveat in the causality violations which could
potentially be achieved using the light, superluminal 
particles: Namely, sending a strong conceivably 
superluminal neutrino burst into a specific direction of space, it
is certainly possible to send a message ``that something happened
somewhere else''
at superluminal speeds, into distant regions of the Universe. 
However, as we show below, it is very difficult to 
reliably ``stamp'' any information onto the superluminal 
neutrinos. Colloquially speaking, the dilemma is this:
High-energy tachyonic neutrinos approach the light cone
and travel only infinitesimally faster than light itself
($E = \sqrt{\vec p^2 c^2 + m^2 c^4} \approx |\vec p|\, c$).
Their interaction cross sections may be sufficiently large to 
allow for good detection efficiency but this is achieved at the 
cost of sacrificing the speed advantage. Low-energy tachyonic 
neutrinos may a substantially faster than light 
[see Eq.~\eqref{Enu}] but their 
interaction cross sections are small and the information 
sent via them may be  lost. 
Even barring arguments which might question the 
practicablity of heavy tachyonic neutrinos for signaling
in principle, it is possible to show that the 
smallness of the cross sections sets important boundaries
for the possibility to transmit information, as follows.

Indeed, turning the argument around and postulating that 
superluminal particles should not have the capacity 
to transport any ``imprinted'' information into the past,
we are naturally led to the assumption that any 
conceivable superluminal particles have to be
very light, and weakly interacting.

We now supplement these considerations with a numerical estimate.
We consider a hypothetical neutrino flavor eigenstate of mass
$m_\nu \leq 1\,{\rm MeV}/c^2$, and assume that the interaction cross
section fulfills \cite{OULU_WEB}
\begin{equation}
\label{oom}
\sigma = A_0 \, \frac{E_\nu}{E_0} \,,
%\qquad
%A_0 \sim 10^{-49} \, {\rm m}^2 \,,
%\qquad
%E_0 = 1\, {\rm MeV} \,.
%\,,
\end{equation}
with $A_0 \sim 1$\,fb and $E_0 = 1$\,GeV, which is satisfied by known
cross sections for charged-current and neutral current tree-level
interactions, as tabulated in Ref.~\cite{OULU_WEB}.  The
order-of-magnitude relation given in Eq.~\eqref{oom} remains valid for
neutrino scattering off electrons, for all three neutrino flavors, even
if additional charged-current interactions exist for electron
neutrinos, due to exchange graphs with virtual $W$ bosons (for muon and
tau neutrinos, only the $Z$ boson contributes at tree level). (The
scale $E_0 = 1\, {\rm GeV}$ is a characteristic scale for the neutrino
energy and has no connection to the neutrino mass $m_\nu$.) A particle
typically cannot be localized to better than an area equal to the
square of its (reduced) Compton wavelength,
\begin{equation}
A_{\rm min} = \lambdabar^2 = \left( \frac{\hbar}{m_\nu \, c} \right)^2 \,.
\end{equation}
The detection probability $P$ for a perfectly focused particle
therefore does not exceed
\begin{equation}
P = \frac{\sigma}{A^2_{\rm min}} = 
\frac{A_0 \, c^4 \, m_\nu^3}{\sqrt{2} \, E_0 \, \hbar^2} \, 
\sqrt{\dfrac{c}{\delta c}} \,.
\end{equation}
If we are to send information reliably, then the detection probability
should be of order unity.  Solving the equation $P=1$ for $\delta c$,
we obtain
\begin{equation}
\label{equal}
\delta c = 
\frac{A_0^2 \, m_\nu^6 \, c^9}{2 \, E_0^2 \, \hbar^4} \approx
10^{-33} \, \frac{\rm m}{\rm s} \,,
\qquad
{\rm for}
\qquad
m_\nu \, c^2 = 1 \, {\rm MeV} \,,
\end{equation}
where we assume a neutrino mass of 1\,MeV$/c^2$.  When traveling at a
speed $c + \delta c$ for a path length $s$, the neutrino acquires a
path length difference of $\delta s$, which compares to its Compton
wavelength as follows,
\begin{equation}
\label{solve}
\delta s = s \, \frac{\delta c}{c} \,,
\qquad
\lambdabar = \frac{\hbar}{m_\nu \, c} \,,
\qquad
\delta s = \lambdabar \Leftrightarrow 
s = \frac{2 \, E_0^2 \, \hbar^5}{A_0^2 \, c^9 \, m_\nu^7} 
\approx 10^{27} \, {\rm m}\,.
\end{equation}
We here use the fact that a path length difference is physically
significant if it exceeds the Compton wavelength of the traveling
particle. So, in conclusion, the distance that the detectable neutrino
($P=1$) has to travel at the small deviation $\delta c$ from the speed
of light given by Eq.~\eqref{equal}, before the accumulated path
difference exceeds its Compton wavelength, is larger than the commonly
assumed size of the Universe, $s_{\rm max} \approx 10^{26} \, {\rm m}$.
This conclusion becomes much stringent for realistic mass values as the
distance $s$ in Eq.~\eqref{solve} is proportional to the seventh
inverse power of the neutrino mass. If we want to prevent information
being sent into the past, then the neutrino mass muss be below $1 \,
{\rm MeV}$, which is satisfied by realistic tachyonic neutrinos.

Finally, we remark that
the permissibility of small violations of the 
``cosmic speed limit'' (the speed of light) and on 
small length and distance scales has been discussed
in the literature previously (see, e.g., Ref.~\cite{AhReSt1998}).
It holds provided quantum effects are properly taken into account.
Furthermore, we refer to the experiments in the 
group of Nimtz~\cite{EnNi1992,NiSt2008,Ni2009}, which also use a compact apparatus
and rely on the quantum mechanical tunneling 
effect, which lies outside the regime of classical mechanics.
As Hilbert put it, ``we must know, we will know''---from experiment.

\end{document}